\documentstyle[aps,epsf,prb,multicol]{revtex}
\begin{document}
\draft
\title{Transport anomaly in the low energy regime of spin chains}
\author{Keiji Saito}
\address{Department of Applied Physics, School of Engineering \\ 
University of Tokyo, Bunkyo-ku, Tokyo 113-8656, Japan}
\date{}
\maketitle

\begin{abstract}
The anomalous thermal conductivity in spin chains 
observed in experiments is studied
for the low temperature regime. In the effective dynamics with most
realistic perturbations, the so-called 
Umklapp terms is irrelevant to reduce mean free path in the 
energy transport at even finite temperatures.
This is consistent with large conductivities
found in recent experiments. 
The Drude weight which is the prefactor in the divergent conductivity
is calculated, and 
the temperature dependence is discussed.
\end{abstract}

\begin{multicols}{2}
\narrowtext
\section{INTRODUCTION}
Recent experiments on transport phenomena in 
low dimensional spin systems have attracted much interest
\cite{SFGOVR00,ATSDTFK98,HLFUKUDR,KINAKMTK01,SGOARBT00,SGOAR00,HBABHBR,KNKNK01,HLUKZFKU01}.
There the  behavior of conductivity is sensitively 
affected by the magnetic phases.  
Therefore the thermal conductivity can be a strong 
experimental instrument to investigate the magnetic properties in
materials. In experiments, magnetic energy transport is much larger 
than the phononic transport at most temperatures.
This fact is considered to be attributed to large
exchange coupling constants and nondiffusive behavior of transport.
Sr$_{2}$CuO$_{3}$ described by the isotropic Heisenberg chain has 
large exchange coupling constant $J\sim 2000K$
and the magnon can transmit energy even at the order of $100$K.
The mean free path is experimentally estimated in the similar way as the 
Peierls-Boltzmann approach for phononic transport \cite{Ktext,C59}, and 
nondiffusive behavior is found. 
In another isotropic Heisenberg chain CuGeO$_{3}$, 
where the mean free path is calculated as $500$ times lattice size. 
It is noteworthy 
that even when the alternation between bonds occurs, the magnetic
transport may be still dominant and nontrivial enhancement of conductivity
is observed \cite{HLFUKUDR,SM02}. 
Spin-ladder system (Sr, La, Ca)$_{14}$Cu$_{24}$O$_{41}$ and 
La$_{5}$Ca$_{9}$Cu$_{24}$O$_{41}$ also show 
extremely large mean free path \cite{KINAKMTK01,SGOARBT00,SGOAR00,HBABHBR}.

As studied in classical systems, an integrable system shows 
ballistic energy transport \cite{RLL67}. 
Even if the energy current is not conserved, the low-dimensional systems
with conserved quantities tend to show the divergence of conductivity
\cite{LLP97}.
These tendencies are also valid in quantum systems 
\cite{Z02,SM02,AG,AG02,MHCB,S02,Huber,Z97}. 
The isotropic Heisenberg chain is the integrable system and exactly 
shows the diverging conductivity as proved by Zotos et al.\cite{Z97}.
This divergence is a direct result of conserved energy current.
However in most materials in experiments, 
the perturbations like a bond-alternation and a next nearest neighbor
interaction exist, so that the energy current is no more conserved. 
Nevertheless the experiments indicate divergent conductivities.
Motivated by these experimental facts, Alvarez and Gros investigated
whether the experimentally indicated divergent
conductivity really occurs or not. They numerically found nonvanishing
Drude weight at finite low temperatures for the alternate spin chain with 
frustration and the ladder system \cite{AG}, although it may vanish
at very high temperatures in the thermodynamic limit\cite{MHCB}. 
Numerical finding of finite Drude weight indicates the existence 
of divergent conductivity at low temperaure regime.
Stimulated by these numerical studies, in this paper we study 
the possible mechanism 
to explain nondiffusive behavior in the presence 
of perturbations to the Heisenberg chain. 
We focus on the low energy regime \cite{note0} 
by employing the effective dynamics as adopted for the Heisenberg chain by 
Meisner et al. \cite{MHCB}.
As a result, we find the effective energy current becomes conserved 
in spite of the existence of so-called Umklapp terms so that the 
conductivity diverges. This ballistic nature is a common 
characteristic for most realistic perturbations 
as far as the low energy regime is considered.
This is a remarkable aspect in the sense that the perturbations
is irrelevant to decrease the mean free path although they 
can be relevant to cause various magnetic phases in the equilibrium state.
This is very important to explain the anomalous large conductivities 
of one-dimensional magnetic systems.
We further calculated the quantitative value of the Drude weight of 
conductivity using the path integral approach, 
and found it to be finite.

\section{EFFECTIVE ENERGY CURRENT IN LOW TEMPERATURE REGIME}
We first consider an alternate spin chain whose Hamiltonian is 
\begin{eqnarray}
{\cal H} &=& {\cal H}_0 + {\cal H}_1 ,\\ 
{\cal H}_0 &=& J \sum_{\ell } {\bf S}_{\ell} \cdot {\bf S}_{\ell +1} , 
\quad {\cal H}_1 = J \delta \sum_{\ell } (- 1)^{\ell +1 }{\bf S}_{\ell} 
\cdot {\bf S}_{\ell +1}.
\end{eqnarray}
The total Hamiltonian is divided into the isotropic part 
${\cal H}_{0}$ and the dimerized part ${\cal H}_{1}$. 
Thermal conductivity $\kappa (T)$ at a temperature $T(=1/\beta)$ 
is calculated by the Green-Kubo formula which reads as \cite{Kubo},
\begin{eqnarray}
\kappa (T) &=&  \lim_{\omega\to 0} \kappa (\omega )
 ={\beta \over 2 L } 
\int_{-\infty}^{\infty} dt e^{-i\omega t}
\int_{0}^{\beta} d\tau \langle j(t-i\tau) j\rangle ,
\end{eqnarray}
where $j (...)$ is the total current operator in the Heisenberg picture.
From the continuity equation of energy, the total current 
operator $j$ is written as 
$j = -i \sum_{\ell} \left[ h_{\ell} , h_{\ell +1} \right]$ with the
local Hamiltonian $h_{\ell}  :=J (1 + (-1)^{\ell +1 }\delta ) 
{\bf S}_{\ell} \cdot {\bf S}_{\ell +1}  $ \cite{SM02}. In the absence of
the alternation, $j$ is the conserved quantity \cite{Z97}, whereas
the existence of alternation does not allow the conservation, i.e.,
\begin{eqnarray}
\begin{array}{ccc}
\left[ j_{(\delta =0)},{\cal H}_{0}  \right] =  0, &{\rm but}&
\left[ j ,{\cal H}  \right] \ne  0.  \label{exact}
\end{array}
\end{eqnarray}

At low energy regime, the Hamiltonian can be effectively 
represented by the boson field \cite{CF79,NF80};
\begin{eqnarray}
{\cal H}_{\rm B} &=& \int \, dx \, 
\left[  \pi v  \, p^2  
+ {v\over 4\pi} (\partial_x \Theta_{+} (x) )^2 
- B \cos \Theta_{+} (x)
\right] \label{alt},
\end{eqnarray}
where $v = {J \pi \over 2}$ and $B= J \delta $. 
The Boson field is defined using the fermion field
$\Psi_{R} $ and $\Psi_{L} $ and satisfies the commutation relation as,
\begin{eqnarray}
{1\over 2\pi} {\partial \Theta_{\pm} \over \partial x} &=& 
: \Psi^{\dagger}_{R} (x)\Psi^{}_{R} (x): \pm 
: \Psi^{\dagger}_{L} (x)\Psi^{}_{L} (x):  ,
\\
p (x) &=& -{1\over 4\pi} \partial_x \Theta_{-} (x) , \\
\left[\Theta_+  (x) , \Theta_-  (x')\right] &=& 2\pi i \,
{\rm sgn} (x-x') .
\end{eqnarray}
We define the local Hamiltonian density at $x$ as,
$h(x) = \pi v p^2 + {v \over 4\pi } 
\left( {\partial \Theta_+ \over \partial x}  \right)^2 
- B \cos \Theta_{+} (x) $.
Using the relations,
\begin{eqnarray}
{\partial \Theta_{+} \over \partial t} &=&  -2\pi v p, \quad 
{\partial p \over \partial t} = -{v \over 2\pi } \partial_{x^2} \Theta_{+} (x)
+B \sin \Theta_{+} ,
\end{eqnarray} 
we find the Heisenberg equation of motion given by,
\begin{eqnarray}
{\partial h(x,t) \over \partial t} &=& -{v^2\over 2} \partial_{x}
\left(  p \, \partial_{x } \Theta_+  + \left( \partial_{x } \Theta_+ \right)
p
\right) 
. \label{heise}
\end{eqnarray}
This leads the effective total energy current written as,
\begin{eqnarray}
j_{\rm B} &=&  v^2 \int dx \, p(x) \partial_x \Theta_{+} (x) . 
\end{eqnarray}
Interestingly, this effective current operator $j_{\rm B}$ does not 
include the perturbation term (Umklapp term) $B \cos \Theta_{+} (x) $. 
Furthermore we find that even if the energy current is not rigorously
conserved (\ref{exact}), it behaves as the conserved quantity in the low
temperature regime, i.e.,
\begin{eqnarray}
\begin{array}{ccc}
\left[ j \, {\cal H} \right] \ne  0, & {\rm but} & 
\left[ j_{\rm B}, \, {\cal H}_{\rm B}\right] = 0. \label{cmt} 
\end{array}
\end{eqnarray}
The perturbation term $B \cos \Theta_{+} (x) $
causes the Umklapp process in the scattering between fermions
\cite{RA00}. Nevertheless
at the level of the energy transport, this process is irrelevant 
to decrease the mean free path at finite low temperatures. 

This situation is also the case in the other realistic perturbations.
We consider the two following examples \cite{H82},
\begin{eqnarray}
{\cal H}_2 &=& J\alpha \sum_{\ell } {\bf S}_{\ell} \cdot {\bf S}_{\ell +2},
\quad {\cal H}_3 = J \gamma \sum_{\ell } (-1)^{\ell} S_{\ell}^{z} ,
\label{other}
\end{eqnarray}    
The perturbation ${\cal H}_2 $ causes the frustration and the nontrivial 
phases appears as a function of $\alpha$ \cite{H82}. Recent numerical 
studies treat this perturbation and the effect of 
frustration is investigated \cite{AG,MHCB}. ${\cal H}_3$ is the term of
staggered magnetic field.
The effective expressions at the low energy regime for the 
perturbations (\ref{other}) are written as,
\begin{eqnarray}
{\cal H}_2  &\sim& \int dx \, c_2
\left(  \partial_{x } \Theta_+  \right)^2 + c_{2}' \cos \left( 2 \Theta_+  \right) ,  \\
{\cal H}_3  &\sim&  c_3 \int dx \, \cos \left( 2 \Theta_+  \right) ,
\end{eqnarray}
with constants, $c_2$, $c_{2}'$ and $c_3$. These cases have the
same form as the Hamiltonian $(\ref{alt})$, and the effective total current is 
easily proved to be conserved. This feature is generally common for 
the Hamiltonian with the form,
\begin{eqnarray}
\int \, dx \, 
\left[ a \, p^2  
+ b (\partial_x \Theta_{+} (x) )^2 
+ \sum_{\ell} c_{\ell} \cos \left( d_{\ell } \Theta_{+} (x) + 
\phi_{\ell} \right)
\right] .
\end{eqnarray}
Here $a,b,c_{\ell},d_{\ell}$, and $\phi_{\ell}$ are constants.
Thus we conclude that the effective energy current is conserved in
most realistic spin chains at least as far as we consider in the low
temperature regime. 

\section{DRUDE WEIGHT}
Now let us calculate the prefactor of thermal
conductivity, i.e., the Drude weight $D_{\rm th}$ in the low energy
regime and confirm that finite value of $D_{\rm th}$ really exists.
Since the energy current does not depend on a
time, we write the Green-Kubo formula as,
\begin{eqnarray}
\kappa &=& D_{\rm th} \delta (\omega)   \\
 D_{\rm th} &=&  
\beta\pi /L 
\int_{0}^{\beta}\, d\tau \int_{0}^{L}\, dx \int_{0}^{L}
d x' \, \langle j_{\rm B} (x, \tau ) j_{\rm B} (x' , 0) \rangle ,
\end{eqnarray}
where $j(x, \tau ) $ is the local energy current at $x$ and the
imaginary time $\tau$ in the Heisenberg picture. The current 
is expressed as $j(x, \tau ) = {-iv\over 2\pi } \, \partial_x \Theta_{+}(x,  \tau ) \, \partial_{\tau } \Theta_{+} (x, \tau)$.
We confine ourselves in the alternate spin chain effectively described
by Hamiltonian (\ref{alt}). At very low temperatures, 
the Hamiltonian (\ref{alt}) is approximated by the mean field theory\cite{DHN74} ; 
\begin{eqnarray}
 \cos \left(\Theta_{+} \right) \sim  e^{-\langle\Theta_{+}^{2} \rangle /2 } 
\left( 1- { \Theta_{+}^{2} - \langle \Theta_{+}^{2} \rangle \over 2  } \right) .
\end{eqnarray}
The Hamiltonian is reduced to,
\begin{eqnarray}
{\cal H}_{\rm B} &\sim & 
\int \, dx \, 
\left[ {v\over 4\pi} (\partial_x \Theta_{+} (x) )^2 
+ C \Theta_{+}^{2}  + \pi v p^2  + {\rm const.}\right]  , \label{appro}
\end{eqnarray}
where $C = B e^{-\langle\Theta_{+}^{2} \rangle /2 } $ and 
$\langle\Theta_{+}^{2} \rangle $ is the average value at the ground
state which is self-consistently determined. This
approximation yields the accurate ground state energy 
within $10$ percent error in comparison with the exact
numerical data \cite{NF80}. Thus we expect that this approximation 
well describes properties in very low temperatures.
Using the Hamiltonian (\ref{appro}),
we calculate the Drude weight $D_{\rm th}$ using the path
integral (e.g.,\cite{S99}). The Drude weight is given by explicit 
calculation of
\begin{eqnarray}
D_{\rm th} &=&
 {-v^2 \over 4\pi \beta L^3 }
\int\int\int d\tau  dx dx' 
\partial_{x_1} \partial_{x_3} \partial_{\tau_2} \partial_{\tau_4}
\sum_{n_1, .., n_4}\sum_{k_1 , .., k_4} \nonumber \\
&\times& e^{i(k_1 x_1 + k_2 x_2 + k_3 x_3 + + k_4 x_4) }
e^{i(\omega_{n_1}  \tau _1 + \omega_{n_2}  \tau _2 + \omega_{n_3}  \tau _3 + 
+ \omega_{n_4} \tau_4  ) } \nonumber \\
&\times& 
 \langle \Theta_{+} (k_1 , n_1 ) 
         \Theta_{+} (k_2 , n_2 ) 
         \Theta_{+} (k_3 , n_3 ) 
         \Theta_{+} (k_4 , n_4 ) \rangle ,
\end{eqnarray}
where we take $x_1 = x_2 =x, x_3 = x_4 = x',\tau_1 = \tau_2 = \tau$, 
and$ \tau_3 = \tau_4 =0$. $k_{m}$ and $\omega_n$ are the wave number 
and the Matsubara frequency defined as $k_m = {2\pi m \over L}$ and 
$\omega_{n} = {2\pi n \over \beta}$, respectively.
$\Theta_{+} (k_m, n )$ is the Fourier transformation at the mode 
$k_m$ and $\omega_n$. 
The four point correlation is calculated by the Wick's theorem using the
two point correlation $ \langle |\Theta_{+} (k_{m} , n )|^2 \rangle 
= {2\pi v \over \omega_{n}^{2}+ v^2 k_{m}^{2} + 4\pi v C }  $.
We next carry out the exact summation over $n_p \,\, (p=1,\cdots ,4)$
using technique of the complex integral.  
We finally obtain the simple expression of the Drude weight with 
$a={\beta\over2\pi}\sqrt{v^2 k^{2} + 4\pi v C}$,
\begin{eqnarray}
D_{\rm th} &=&   {v^4 \beta^2 \over 8 }
\int_{-\infty}^{\infty} dk k^2 \left[ \left( 
      { 1 \over e^{2\pi a } -1 } 
    + { 1  \over e^{ - 2\pi a } -1 } 
\right)^2  \right. \nonumber \\ 
&& \hspace*{2.5cm} - \left. \left( 
      { 1 \over e^{2\pi a } -1 } 
    - { 1   \over e^{ - 2\pi a } -1 } 
\right)^2 
\right]  \nonumber \\
&=& 2v T \int_{0}^{\infty} dx \, 
{x^2  \over \sinh^2 \sqrt{ x^2 + \pi \beta^2 v C} } . \label{result}
\end{eqnarray}
In the case of isotropic Heisenberg chain $C=0$, 
$D_{\rm th} = {v\pi^2 T\over 3}$ because of $\int_{0}^{\infty} dx \, 
{x^2  \over \sinh^2 x } = {\pi^2 \over 6} $, which 
completely agrees with the asymptotic behavior of exact calculation by 
the Bethe ansatz \cite{KS01}. This result is also given
by the conformal transformation \cite{MHCB,note2}.
In the presence of finite alternation $\delta$, however,
the Hamiltonian does not have conformal symmetry. 

Eq. (\ref{result}) shows that the Drude weight 
is non-vanishing for finite alternation $\delta$.
This is quite crucial because 
the existence of $D_{\rm th}$ assures the anomalous energy transport.
The energy gap $\Delta E$ is roughly estimated as 
$\Delta E \sim 2\sqrt{\pi v C}$. Thus for very low temperature regime 
$T \ll \Delta E$, eq.(\ref{result}) shows
exponential decrease with power law prefactor due to the finite gap 
\begin{eqnarray}
D_{\rm th} &\sim &{v  \beta^2 \over 32} \int_{-\infty}^{\infty}
dp \, p^2 e^{-\beta \left( \Delta E + {p^2 \over 2 \Delta E}\right)} \nonumber
\\
&=& {v\over 32}\sqrt{\pi \Delta E^3 \over T} 
\, e^{-\beta \, \Delta E} \qquad 
 T \ll \Delta E. \label{lowtemp}
\end{eqnarray}
Exponential decrease is simply due to exponential decrease of thermal 
excitation.   
The power law prefactor $T^{-1/2}$ is also seen in 
the electric and spin conductivity in one-dimensional gapped systems 
\cite{DS,GR}. This temperature dependence will be the characteristic 
in transport phenomena in the low-dimensional gapped systems.

\section{SUMMARY}
We find the possible 
mechanism of anomalous transport in low energy
region of realistic spin chains. In the isotropic Heisenberg chain, 
the energy current is exactly conserved so that the Green-Kubo 
formula trivially diverges. On the other hand, the presence 
of perturbations, the Umklapp terms appear. However this term is 
irrelevant to the energy current of long wave length at 
low temperatures.
We considered the quantitative estimation of the Drude weight for the
alternate chain, and find exponential decrease of Drude weight at low 
temperature. This aspect is consistent with the numerical studies 
\cite{AG,MHCB}.

Here we should make clear the difference between classical nonintegrable 
Fermi-Pasta-Ulam (FPU) chains \cite{LLP97} where the mechanism of
anomalous transport is still argued.
In this classical system, anomalous heat transport is attributed to
a slow relaxation due to the total momentum conservation, whereas
the the auto-correlation function
of energy current vanishes in the long time limit, i.e.,
the mixing property is satisfied.
However, in perturbed quantum Heisenberg spin chains we discuss here, 
Numerical evidence of the finite Drude weight \cite{AG}
means the violation of the mixing property. 
Thus the mechanism in quantum spin chains is different from the one in
classical FPU system.

The alternate spin chain directly related to 
the spin-Peierls system. In the realistic mechanism, the alternation is
formed by the fluctuation of the phonon. In this sense, in more realistic 
treatment, $\delta$ must be exchanged by the time-dependent 
bond length between atoms. 
In the case where the bond fluctuation is not negligible, 
how the magnetic energy transport is affected is also interesting. \\
\noindent
{\em note added}--
After submitting this paper, we find the preprint \cite{OCC02} where 
the Drude weight in the dimerized xy spin chain and two-leg ladder system
are extensively studied. The paper includes the similar formula 
as (\ref{result}) in these systems.

\section*{Acknowledgments}
The author would like to thank S. Miyashita and X. Zotos for useful comments.
He also expresses thanks the referees for suggestions to interpret the 
formula (\ref{result}).
The present work is supported by Grand-in-Aid for Scientific 
Research from Ministry of Education, Culture, Sports, Science,
and Technology of Japan.

\end{multicols}
\end{document}